\magnification=1200
\vsize=8.5truein
\hsize=6truein
\baselineskip=20pt
\centerline{\bf High genus periodic gyroid surfaces}
\centerline{\bf of non-positive Gaussian curvature}
\centerline{by}
\centerline{Wojciech G\`o\`zd\`z and Robert Ho{\l}yst}
\vskip 20pt
\centerline {Institute of Physical Chemistry
PAS and College of Sciences,}
\centerline {Dept.III, Kasprzaka 44/52,
01224 Warsaw, Poland}

\centerline{\bf Abstract}
In this paper we present the novel method for the generation of
periodic embedded surfaces of nonpositive Gaussian curvature.
The structures are related to the local minima of the
scalar order parameter Landau-Ginzburg hamiltonan for microemulsions.
The method is used to generate six unknown surfaces of Ia$\bar 3$d
symmetry (gyroid)
of genus 21, 53, 69, 109, 141 and 157 per unit cell.  All of them
but that of
genus 21  are most likely the minimal surfaces.
Schoen-Luzzati gyroid minimal surface of genus 5 (per unit cell)
is also obtained.

\vfill\eject

The surfactant molecules, which are the
main ingredient of soaps and detergents have the ability to solubilize
oil in water, two liquids which in the binary mixture at normal conditions
are immiscible. This ability stems from their chemical structure;
a surfactant molecule  has the
polar and nonpolar segments at two ends and thus is simultaneously
hydrophobic and hydrophilic. The term amphiphilic
(from Greek word: loving both)
molecule is used
since one end (polar) of the molecule is well solubilized in water while the
other (nonpolar) in oil. Hence the molecule
preferably stays at the oil-water interface,
 forming a monolayer.
At high concentration of surfactant the physical interface
made of these molecules orders, forming periodic structures of various
symmetries.
Similar behavior is observed in systems of
biological molecules (lipids) which in water solutions
self assembly into bilayers.
In 1967/68 Luzzati et al$^{1-3}$
observed that the type of ordering
in the lecithin-water and lipid-water systems cannot be described by the
arrangement of simple surfactant aggregates such as cylinders, planes or
spheres. They observed the cubic bicontinuous phase of the Ia$\bar 3$d
symmetry (gyroid) where the lipid bilayers formed a highly curved smooth
(embedded) surface of the same symmetry. Such surface divides the
volume into two disjoint subvolumes.
The NMR, SAXS (small angle x-ray scattering) and surfactant concentraction
measurements indicate that these surfaces closely
resemble triply periodic minimal surfaces$^{4-7}$ i.e.
surfaces characterized by  zero mean curvature at every
point. The latter belongs to the broader
class of periodic surfaces of nonpositive Gaussian curvature.
Since the discovery of minimal periodic surfaces
in 1865 by Schwarz only one periodic
embedded gyroid surface of cubic
symmetry and  genus 5 (per unit cell)
has been discovered and fully characterized$^4$.
Apart from that six more of Pn$\bar 3$m and
Im$\bar 3$m symmetry are known, all of low genus.
{\it Here we present the general
method which can be used to generate periodic surfaces of nonpositive Gaussian
curvature. We prove the efficiency of this method by generating
six new gyroid structures of genus 21, 53, 69, 109, 141 and 157.
All these surfaces but that of genus 21  are minimal;
we have also generated the
Schoen-Luzzati gyroid minimal surface of genus 5 per unit cell.
}

We point out that the unique characterization of the periodic surfaces
using standard x-ray scattering methods is very difficult if the
motif in the unit cell is not precisely known$^{8-11,12}$. Our theoretical
method combined with experimental technique will be extremely helpful in
future discoveries of new surfaces in biological systems.

The surfaces of
surfactant systems find application in the production of the mesoporous
silicate systems, where in the synthesis process the ordered surfactant surface
is used as a template for the three dimensional polimerization of silicate$^
{13,14}$.
One obtains an ordered silicate pore system with the symmetry and geometry
of the surfactant template.
We believe that our method can also be
used for the design of new mesoporous structures.

Our method is based on the
Landau-Ginzburg model which has been
proposed by Teubner and Strey$^{15}$ and Gompper and
Schick$^{16,17}$
on the  basis of neutron scattering experiments performed on
microemulsion  (homogeneous ternary
mixture  of oil, water and surfactant) and later experiments and
theory of their wetting properties$^{18,19}$.
The Landau-Ginzburg free energy functional
has the following form:
$$F[\phi]=\int d^3 r \left(\left\vert\triangle \phi\right\vert^2+
g(\phi)
\left\vert\nabla \phi\right\vert^2+(\phi^2-1)^2(\phi^2+f_0)
\right)\eqno(1)$$
where
$g(\phi)=g_2\phi^{2}-g_0.$
Here $\phi$, the order parameter, has the interpretation of the
normalized
difference between oil and water concentrations; $g_2$, $g_0$ are
positive constants and $f_0$ can be of either sign.
The  last term in Eq.(1) is the bulk free energy and describes the
relative stability of the pure water phase ($\phi=-1$), pure oil phase
($\phi=1$)
and microemulsion ($\phi=0$).
The
stability of bulk microemulsion phase
depends on $f_0$:
for $f_0>0$ microemulsion is a metastable bulk phase
whereas pure water phase
or pure oil phase are stable; for $f_0\le 0$ microemulsion is
stable.
We note that in
general $g(\phi)$ can be a polynomial in $\phi^2$.

For $g_0>2$
the system can undergo a transition to periodically ordered phases
where water rich domains and oil rich domains order. The interface between the
domains corresponds to $\phi ({\bf r})=0$. The different structures
(stable or metastable) correspond to the minima of the functional (1).
The following simple argument shows that among the surfaces,
inside these structures, we might expect minimal surfaces.
The mean curvature of the surface at point {\bf r} is
given by the divergence of the vector normal to the surface at this point
$^{20,21}$:
$$H=-{1\over 2}
\nabla\left({{\nabla\phi}\over{\vert\nabla\phi\vert}}\right)=
-{1\over 2}{{\triangle\phi}\over{\vert\nabla\phi\vert}}+
{{\nabla_n\vert\nabla\phi
\vert}\over{2\vert\nabla\phi\vert}}.\eqno(2)$$
Here $\nabla_n$ denotes the derivative along the normal to the surface.
It follows
from the second term of Eq(1) that
$F[\phi]$ is minimized when  $\vert\nabla\phi\vert$ has the
maximal value  for
$\phi({\bf r})=0$ since at that point $g(\phi)$ has the lowest
value. For the maximum of $\vert\nabla\phi\vert$
its normal derivative vanishes and consequently  the
second term in Eq(2) does so. We also know that in the case of
$\phi$, $-\phi$ symmetry, $H$ averaged over the whole
surface should be zero. It means that either $\triangle\phi$ is exactly zero
at the surface or it changes sign. From the first term of Eq(1) it
follows that the former can be  favored and consequently $H=0$ at every
point at the surface. Hence we can expect that some of the
surfaces are minimal. This argument
does not take into account the global distribution of the field $\phi$,
nonetheless it provides a useful hint for our studies.

In order to find the minima of the
functional we have discretized Eq(1)
on the cubic lattice.
Thus the functional $F[\phi({\bf r})]$ becomes a function
$F(\{\phi_{i,j,k}\})$ of $N^3$
variables, where $Nh$ is the linear dimension of
the cubic lattice and $h$ is the distance between the lattice points.
Each variable $\phi_{i,j,k}$ represents
the value of the field $\phi({\bf r})$ at the lattice site $(i,j,k)$,
and the indices $i,j,k$ change from 1 to N. In our calculations we use
N=17,33,65 and 129; final results are shown for N=129.
Please note that N=129
results in over 2 milion points per unit cell.  The first and
second derivatives in the gradient and laplasian terms of the functional (1)
were calculated on the lattice according to the three point formula for
the first derivatives and five point formula for the second derivatives
$^{22}$.
We impose on the field $\phi_{ijk}$  the periodic boundary
conditions and the symmetry of the structure,
we are looking for, by building up the field inside a unit
cubic cell from a smaller polyhedron, replicating it by
reflections and rotations combined with translations, since the gyroid
symmetry involves glide planes.
Such procedure enables substantial
reduction of independent variables.

The initial configuration needed for the minimization
is set up by building the field
$\phi ({\bf r})$ first on a small lattice $N=3$ or 5. It is done
by analogy to the structure of
a two component (A,B) molecular crystal. The value of the field
$\phi_{i,j,k}$ at a lattice site $(i,j,k)$ is set to $1$ if in
the molecular crystal an atom A is in this place. It is set to
$-1$, if there is
an atom B, and to $0$, if there is an empty place
at $(i,j,k)$. Next the small lattice can be enlarged
to desired size by changing the number of points from
$N$ to $2N-1$ and finding the values of $\phi_{i,j,k}$
in new lattice sites by interpolation.

We have used the
conjugate gradient method$^{23}$ to find a minimum of the function
$F(\{\phi_{i,j,k}\})$.
It is highly unlikely, because of numerical accuracy,
that a value of the field $\phi_{i,j,k}$
at a latice site $(i,j,k)$ is exactly zero.
Therefore the points of
the surface have to be localized by linear interpolation between the neighbour
sites of the lattice. This approximation is legible
because the field $\phi({\bf r})$ is very smooth.
The points on the surface are used for the triangulation. From the
triangles covering the surface we get the surface area and the Euler
characteristic,
$\chi$. The latter is given by the Euler relation:
$\chi=F-E+V$,
where $F$ is the number of faces, $E$ is the number of edges and $V$ is the
number of vertices of the triangles covering the surface.
The edges and vertices has to
be taken with weight  $1$, $1/2$, or $1/4$ if they appear inside, at the face,
 or at the edge
of the unit cell, respectively.

We have performed the detailed study of the phase diagram, checking the
Landau free energy of almost {\it 30 different structures of various
symmetries}
and found that
the only
stable ordered structure is the
lamellar phase. The phase boundaries for the lamellar phase are given by
Gompper and Zschocke$^{24}$.
The gyroid phase of genus 5 has the second lowest energy
(after lamellar phase) among all the structures (Table 1).
We have noted that this phase has larger area per unit volume than the lamellar
phase, thus in the case of very sharp interfaces it should have smaller energy
than the lamellar phase, since from the second term of
Eq(1) it follows that at $\phi({\bf r})=0$ the gradient term gives large and
negative contribution to the energy. This negative contribution is
not cancelled by the bulk or laplacian terms.
Unfortunately the size of the interface scales with
the size of the unit cell.  We have performed the same calculations for
the new function $g(\phi )$ given by :
$g(\phi )=g_2\phi^4-g_0.$
Increasing the power of $\phi$ by a factor of two indeed sharpens the
interface between oil and water
but at the same time reduces the size of the unit cell.
The net result is the larger relative difference in energies
between the gyroid and lamellar phase. We note that
in the case of
multiparameter Landau models introduced in recent years$^{16,25}$
we may expect the stabilization of the various phases which
here are only metastable.

Among the local minima of the functional (Eq(1)) we have found
four known minimal surfaces: P, D, I-WP and G$^{26}$.
Here we present the detailed
study of the surfaces of gyroid symmetry.
In all these structures a surface is characterized by non-positive (zero or
negative)
Gaussian curvature. In Table 1 the main characteristics of the gyroid surfaces
are given. Here we have used $N=129$ points per edge of the unit cell.
In order to estimate the errors we have compared the results
obtained for $N=65$ and $N=129$; for the surface area the largest errors
(gyroid 141 and 157) are smaller than 0.3\%, for the energy the
largest error is smaller than 1\% for the gyroid 5 structure and
few percent for other structures. Of course these are the upper limits
and most probably the errors are much smaller.
In Figs.(1,2) two gyroid structures are shown
together with the histograms of their mean curvature.
Please note that for the minimal surface
the mean curvature is peaked around 0 in the histograms, but
due to the numerical accuracy the peak has a finite width.
 We note that as the genus of the surface increases the surface
area per unit volume ($S/d^3$) and the energy per unit volume does not change
very much(Table 1).
Some authors$^{27}$ ruled out the possibility of existance of
high genus surfaces in real systems, because of expected high curvature
regions. We have checked the Gaussian curvature in high genus surfaces and
found that it is not much different from the
low genus surfaces. This is
due to the sufficiently large size of the unit cell
for the former structures.
We also observe that the gyroid structures of high genus
are most easily generated (from any initial
configuration and sufficiently large
unit cell)
close to the stability region of microemulsion.

Summarizing: We have used the Landau-Ginzburg hamiltonian for microemulsions
to generate periodic surfaces of nonpositive Gaussian curvature.
Using this model we have studied 7 periodic gyroid surfaces: one of them
is the Schoen-Luzzati minimal gyroid surface of genus 5. The
remaining six structures are new.
The model can be well applied by physicists working in soft
condensed matter, mathematicians working in topology, biologists and
crystallographers. We are positive that its richness is far
from being explored by our work.

This work was supported by the KBN grants 2P03B 01810 and 30302007
and Foundation FWPN.
\vfill\eject
\centerline{\bf References}
\item{1.} V. Luzzati, T. Gulik-Krzywicki, A. Tardieu,  Nature
{\bf 218}, 1031 (1968).
\item{2.} V. Luzzati, P.A. Spegt,  Nature {\bf 215}, 701 (1967).
\item{3.} V. Luzzati, A. Tardieu, T. Gulik-Krzywicki,
 Nature {\bf 217}, 1028 (1968).
\item{4.} A.H. Schoen,   NASA Technical Note D-5541,
Washington D.C., USA (1970).
\item{5.} D.M. Anderson, H.T. Davis, J.C.C. Nitsche, L.E. Scriven,
 Adv.Chem.Phys. {\bf 77}, 337 (1990).
\item{6.} A.L. Mackay, Nature {\bf 314}, 604 (1985).
\item{7.} H. Terrones, J. de Physique Colloque {\bf 51}, C7 345
(1990).
\item{8.} J.M.Seddon and R.H.Templer, Phil.Trans. R.Soc.Lond. {\bf 344}, 377
(1993).
\item{9.} P.Mariani, L.Q. Amaral, L.Saturni and H. Delacroix , J.Phys. II
France
{\bf 4}, 1393 (1994).
\item{10.} M. Clerc and E. Dubois-Violette, J.Phys II France {\bf 4},
275 (1994).
\item{11.} T. Landh, J.Phys.Chem. {\bf 98}, 8453 (1994).
\item{12.} L.E.Scriven Nature {\bf 263}, 123 (1976).
\item{13.} C.T. Kresge, M.E. Leonowicz, W.J.Roth, J.C.Vartuli and J.S.Beck
Nature {\bf 359}, 710 (1992).
\item{14.} A. Monnier et al,  Science {\bf 261}, 1299 (1993).
\item{15.} M. Teubner, and R. Strey, J.Chem.Phys. {\bf 87}, 3195 (1987).
\item{16.} G.Gompper and M.Schick, {\it Self Assembling Amphiphilic Systems},
vol. {\bf 16} {\it Phase Transitions and Critical Phenomena} eds. C.Domb and
J.L.Lebowitz, Academic Press (1994).
\item{17.} G. Gompper and M. Schick, Phys.Rev.Lett. {\bf 65}, 1116
(1990).
\item{18.} K.-V. Schubert and R. Strey, J.Chem.Phys. {\bf 95},
8532 (1991).
\item{19.} G. Gompper, R. Ho\l yst and M. Schick, Phys.Rev. A
{\bf 43}, 3157 (1991); J. Putz, R. Ho\l yst and M. Schick,
{\it ibid} {\bf 46}, 3369 (1992); Phys.Rev. E {\bf 48}, 635 (1993).
\item{20.} I.S. Barnes, S.T. Hyde and B.W. Ninham, J. de Physique Colloque
{\bf 51} C7, 19 (1990).
\item{21.} M. Spivak, M. {\it A Comprehensive Introduction to Differential
Geometry} vol {\bf III } Publish or Perish Berkley 202-204 (1979).
\item{22.} {\it Handbook of Mathematical Functions With Formulas, Graphs and
Mathematical Tables} eds. Abramowitz and Stegun, I.A., NBS Applied Mathematics
Series
{\bf 55} 883-885 (1964).
\item{23.} W.H., Flannery, B.P., Teukolsky, S.A., and Vetterling,
W.T., {\it Nu\-me\-ri\-cal Re\-ci\-pes} , pp 301- 307 , (1989)
\item{24.} G.Gompper and S.Zschocke, Phys.Rev.A {\bf 46}, 4836 (1992).
\item{25.} A.Ciach, Polish.J.Chem. {\bf 66}, 1347 (1992).
\item{26.} W.G\'o\'zd\'z and R.Ho\l yst, Macromol. Theory Simul. (in press)
(1996).
\item{27.} S.T.Hyde, J.Phys.Chem. {\bf 93}, 1458 (1989).
\item{28.} G.Gompper and M.Kraus, Phys.Rev. E {\bf 47}, 4301 (1993).

\vfill\eject
\centerline{\bf Table Caption}
\item{Table 1.} The gyroid surfaces of nonpositive Gaussian
curvature.
The symmetry of all the structures is Ia$\bar 3$d. In all cases the
volume fraction is 0.5 by construction. Here $g_2=4\sqrt{1+f_0}+g_0+0.01$,
 $g_0=3$ and $f_0=0$. At this point the energy of the stable
phase (lamellar phase) is -0.2077 and the size of the unit cell is $d=3.4$.
In column 2 the energy per unit volume is given. In the third column
the dimensionless linear size of the unit cell, $d$ is given.
The surface area, $S$, (fourth column) is divided by $V^{2/3}=d^2$ i.e. is
calculated per face of the unit cubic cell. The surface area per unit volume,
$S/d^3$, is almost constant for all the structures. The genus, $g$,  (fifth
column)
has been calculated from
the formula $1-\chi/2$, where $\chi$ is the Euler characteristic per unit cell.
We give, in the last column, the
quantity $\delta=\vert\chi\vert^{1/3}d^2/S$. This quantity characterizes not
only
the ordered phase, but also the fluctuating microemulsion$^{28}$.
We think that it can be used as a test for the structure
of microemulsion.
We
find these structures practically for all values of the parameters where
the lamellar phase is also stable$^{24}$, although most easily they are
generated
close to the microemulsion stability region. The genus,
surface area per side of the
unit cell $S/d^2$, symmetry, volume fraction and $\delta$ do not depend on the
parameters, $g_0$,$f_0$ and $g_2$. Only the energy, size of the unit  cell
and  surface area per unit volume are model dependent.

\centerline{\bf Figure Captions}

\item{Fig.1} G141 gyroid structure (see Table 1). {\bf (a)}
One unit cell {\bf (b)} The
histogram of the mean curvature $H$. The size of the unit cell is given in
Table1.
Herethe typical curvature, $1/R=\pm
\sqrt{-K}$, is
$\pm 0.1$ ($K$ is the Gaussian curvature).
\item{Fig.2} GM157 gyroid structure. Legend as in Fig.1.
Here the typical curvature is $\pm 0.5$.

\vfill\eject\end